# Goals and Strategies for the Indexing of Publication Types and Study Designs

Neil R. Smalheiser[1], MD, PhD, Joe D. Menke[2], MS, Arthur W. Holt[1], MBA, Halil Kilicoglu[2], PhD, and Jodi Schneider[3], MA, MSLIS, PhD

[1] Department of Psychiatry, College of Medicine, University of Illinois at Chicago, Chicago, Illinois, USA
[2] School of Information Sciences, University of Illinois Urbana-Champaign, Champaign, Illinois, USA
[3] Information School, University of Wisconsin-Madison, Madison, Wisconsin, USA

Corresponding Author: Neil R. Smalheiser, Department of Psychiatry, College of Medicine, University of Illinois at Chicago, Chicago, Illinois, USA 60612; neils@uic.edu. 708-369-3591

## ABSTRACT

**Objectives.** Major research and implementation efforts have been devoted to indexing articles according to the major topics discussed, but much less effort to indexing their publication types and study designs (collectively, PTs). In this Perspective, we discuss how indexing PTs differs from topical MeSH indexing and requires a different approach.
**Materials and Methods.** Rather than focus on the technical aspects of machine learning-based indexing models, we emphasize the goals and purposes for which biomedical articles are indexed, and the surprisingly thorny question of how indexing systems should be evaluated.
**Results.** Topical Medical Subject Heading (MeSH) terms are assigned to articles that cover the major topics discussed; when more than one term is applicable, only the most specific term is assigned. In contrast, PTs are assigned to articles that have a given structure or use a particular design. To meet the needs of end-users, particularly groups involved in evidence syntheses, PT indexing needs to be comprehensive and employ probabilistic goodness-of-fit prediction scores. Whereas existing NLM hierarchies place publication types and study design-related terms on separate trees from each other, we have created a unified hierarchy that permits more appropriate retrieval via automatic expansion.
**Discussion.** Automated PT indexing systems should allow users to input article records or full-text PDFs and receive scores in real time. This will offer consistent indexing across bibliographic databases, as well as preprints and unpublished manuscripts.
**Conclusions.** Automated PT indexing systems, properly designed and implemented, hold the promise of greatly improving the retrieval of biomedical articles, saving substantial effort when writing evidence syntheses and benefiting other users as well.

**Database URL:** https://www.nlm.nih.gov/medline/medline_home.html

## INTRODUCTION

The biomedical informatics community has conducted major research and implementation efforts for indexing articles according to the major topics discussed (e.g., 1-4). In part, this is intended to overcome ambiguities and limitations of using keywords for online queries; for example, the word "neural" could refer either to neural pathways or to neural networks. Indexing according to the most important ~8-20 topics discussed ensures that users will not retrieve articles in which a given term is mentioned incidentally. As well, establishing a hierarchy of Medical Subject Heading (MeSH) terms permits automatic expansion of search terms to ensure better recall; for example, a search on "central nervous system" will also automatically retrieve articles dealing with more specific terms such as "brain" and "spinal cord".

Medical Subject Heading (MeSH) terms are arranged hierarchically in 16 different trees (Table 1). Whereas most of the trees contain topical MeSH terms that describe entities discussed in biomedical articles (e.g., Humerus, *C. elegans*, India, etc.), the Publication Characteristics tree describes the features of the articles themselves, and will be referred to as the Publication Types (PT) hierarchy (https://www.nlm.nih.gov/mesh/pubtypes.html). This tree includes a wide variety of general article types (e.g. Biography, Letter, Editorial, Scientific Integrity Review) as well as medical types (e.g., Case Reports, Clinical Trial, Practice Guideline, etc.).

**Table 1. The MeSH Hierarchy showing high-level trees.**

- Anatomy [A]
- Organisms [B]
- Diseases [C]
- Chemicals and Drugs [D]
- Analytical, Diagnostic and Therapeutic Techniques, and Equipment [E]
- Psychiatry and Psychology [F]
- Phenomena and Processes [G]
- Disciplines and Occupations [H]
- Anthropology, Education, Sociology, and Social Phenomena [I]
- Technology, Industry, and Agriculture [J]
- Humanities [K]
- Information Science [L]
- Named Groups [M]
- Health Care [N]
- **Publication Characteristics** [V]
    - Publication Components [V01]
    - Publication Formats [V02]
    - Study Characteristics [V03]
    - Support of Research [V04]
- Geographicals [Z]

The Publication Characteristics tree (bold) is expanded to show its leading sub-trees.

According to a National Library of Medicine (NLM) website: "MeSH indexing for MEDLINE was done completely by human indexers until 2011. At that point, while the vast majority of journals were still done by full human indexing, NLM began experimenting with a method of indexing that, for a select set of journals, involved human curation of algorithmic results. The algorithm involved was the Medical Text Indexer (MTI), which was developed by researchers in the Lister Hill Center, and this method of indexing was termed MTIFL (MTI-First Line). In 2019, they began experimenting with the full automation of indexing for some journals, with human review of select citations. The algorithm involved was a customized version of MTI termed MTIA (MTI-Auto). In 2024, they released the MTIX (Medical Text Indexer-NeXt Generation) algorithm (https://www.nlm.nih.gov/pubs/techbull/ma24/ma24_mtix.html), which uses a deep neural network model" (https://www.nlm.nih.gov/bsd/indexfaq.html#produced).

In general, when a term is placed within a hierarchy of more general and more specific terms, only the most specific term is chosen for a given article. That is, if one looks at the MEDLINE XML metadata for that article, only the specific term is present, even though logically the more general terms should apply as well. The XML metadata for an article also indicates whether the MeSH terms have been 1) assigned automatically with human supervision (indexingmethod_curated); 2) entirely automatically (indexingmethod_automated), or 3) assigned manually (medline[sb] NOT (indexingmethod_curated OR indexingmethod_automated).

Most of the research effort in developing automated MeSH indexers has focused on topical MeSH terms. Although automated MeSH indexing systems have been used to index terms in the Publication Characteristics tree as well (5), much less effort has been devoted to the specific issues involved in indexing articles according to the subset of MeSH terms that encompass their publication types and study designs (collectively referred to here as PTs).

Indeed, there is a fundamental difference in strategy between assigning the majority of MeSH terms to an article, and assigning one or more PT terms. Namely, topical MeSH terms are generally assigned to the major topics that are DISCUSSED in an article. In contrast, one or more of the Publication Types (classified under Publication Formats and Study Characteristics) are assigned to an article if it actually HAS a given structure or uses a particular design. To complicate matters, there are also two situations in which a topical MeSH term may refer to a publication type: One is “as topic” MeSH terms, for example, “Clinical Trials as Topic”, which is assigned when an article discusses one or more clinical trials or more general aspects of trial methodology or design. More importantly, a subset of MeSH terms refers directly to study designs, for example, Cohort Studies, Case-Control Studies or Cross-Sectional Studies. These are indexed by NLM according to two different strategies at once, i.e. they may be assigned to articles that DISCUSS the given design, but may also be assigned to articles that EXHIBIT the given design. For example, NLM assigns the MeSH term "Cross-Sectional Studies" both to systematic reviews that include this type of study (perhaps among others) in their review, as well as to observational studies that employ a cross-sectional design. This creates an ambiguity for users (particularly those engaged in evidence synthesis) hoping to selectively retrieve articles that actually employ that design.

In this Perspective, we will demonstrate that indexing PTs, that satisfy (not merely discuss) both Publication Types and study designs, has unique goals and technical considerations that differ from topical MeSH indexing, and that require taking a different approach.

**PREDICTIVE FEATURES**

Not only are the goals of indexing topical MeSH terms vs. PTs different, but the features and architecture of the respective automated indexing models are quite different too. For example, when seeking to index articles on (say) Calcium Channels, the most important predictive features will include topical text words related to channels directly (e.g. "voltage-dependent") but also drugs that interact with these channels, diseases associated with these channels, and other words related to disciplines which study calcium channels (e.g., physiology, cardiovascular). In contrast, one would like to identify (say) a Case-Control Study regardless of whether it studies grip strength in dementia patients, wound healing in children, or tumor treatments in dogs. Thus, topical text terms may be irrelevant and possibly even detrimental to a predictive PT model (6).

**COMPREHENSIVENESS**

Automated MeSH indexing systems currently assign a list of ~8-20 most important MeSH terms to a given article. The list may or may not be explicitly ranked in terms of importance, but only a small subset of all terms discussed in the paper are indexed, and there is no guarantee that all applicable PTs and study designs will be included in the list.

In contrast, when indexing PTs and study designs, all terms that are satisfied should be assigned comprehensively, regardless of the total number or perceived importance relative to other terms. This is important for medical librarians in order to have the biomedical literature classified accurately. But a far more important reason is to facilitate the comprehensive retrieval of articles having a given design, for use in collecting evidence to be analyzed in evidence syntheses (i.e., meta-analyses, systematic reviews, scoping reviews, etc.). Such syntheses generally seek to consider all available evidence, but only from certain specific designs. In practice, the NLM indexing scheme is not complete nor reliable enough to meet these needs (7-9). As a result, evidence synthesis groups typically carry out large initial searches that retrieve thousands of articles, followed by extensive manual screening of titles and abstracts (using at least two independent screeners), to obtain perhaps a few dozen articles for further full-text examination. In other words, because current PT indexing is inadequate for evidence synthesis workflows it is therefore effectively ignored.

In efforts to mitigate this problem, much of the recent work has focused on developing LLM-based methods for title and abstract screening (10-13). While valuable, these approaches primarily address downstream screening burden rather than the upstream indexing problems. From this perspective, the rapid growth of LLM-based screening methods can be seen as compensating for, rather than correcting, deficiencies in the initial retrieval step, providing a strong motivation for a comprehensive automated PT indexing strategy.

## PROBABILISTIC GOODNESS-OF-FIT SCORING

Topical MeSH terms generally refer to concrete entities that have clear definitions: the Humerus is a specific bone in the arm, *C. elegans* is a species with clear genetic and phenotypic features, and India is a specific country with definite boundaries (at a given point in time). In contrast, Publication Types are abstract concepts, and many have neither rigid definitions nor exclusive classifications (see also (7)):

For example, in the biomedical literature, clinical Case Reports are unplanned eyewitness reports of one or a few patients, in contrast to case series which are generally planned analyses (e.g., retrospective chart review) of a larger set of patients, generally four or more and sometimes hundreds (14). However, the MeSH database defines Case Reports as "Clinical presentations that may be followed by evaluative studies that eventually lead to a diagnosis," a definition that does not distinguish typical case reports from typical case series at all (https://www.ncbi.nlm.nih.gov/mesh/68002363), and indeed, many articles that authors call "case series" are indexed by NLM as Case Reports (15). Thus, an article may be a lesser or better fit to the canonical description of a case report. Many case series studies also fit definitions of other publication types, e.g., population cohort studies (16, 17).

Thus, we have argued that instead of simply scoring an article yes/no according to whether it fits a formal definition or not, it is more appropriate to give a predictive score to an article indicating how typical it is of the other articles indexed with that PT. One might consider this as a measure of goodness-of-fit or likelihood. However, we prefer to express the predictive score as a number $0 \leq p \leq 1$, expressed as the probability that a curator would assign the given PT to that article (18).

Articles lacking the design entirely will receive scores near zero; canonical examples of the study design will receive scores near 1; and those with atypical, mixed or unclear features may receive scores near 0.5. Thus, for 100 articles that have a predictive score of 0.5 for (say) Randomized Controlled Trial, this score predicts that a curator would assign that tag to 50 of those articles (18). For articles that comprise multiple publication types (e.g., many case reports include a review of the literature), each PT is considered and scored independently.

To give another example, a "typical" Clinical Trial offers a treatment intervention to patients with a particular disease and looks for changes in health outcomes. Yet trials comprise a wide variety of designs, including studies in which hospitals or schools are the unit rather than an individual patient; studies where the interventions may be diet, exercise or meditation; and outcomes that may be subjective or reflect activities of daily living. A partial solution is dividing "clinical trial" into different design subtypes (e.g. adaptive, pragmatic, phase I, etc.) each scored as a separate PT. However, even so, a given clinical trial article may be more or less typical of its category.

Finally, let us consider "The Autobiography of Alice B. Toklas" by Gertrude Stein (19). Is this an autobiography, a biography, or both? One can make an equally compelling case for all three decisions. An autobiography (by definition) is a person writing about themselves, whereas the Cambridge Dictionary definition of a biography is the life story of a person written by someone else (https://dictionary.cambridge.org/dictionary/english/biography). Clearly the book is a biography and not an autobiography (since Gertrude Stein is not Alice B. Toklas), yet it has the structure/design of an autobiography. If one seeks to index a document according to its design, it should be deemed an autobiography, not a biography. In fact, the NLM hierarchy considers Autobiography to be a subset of Biography (https://www.ncbi.nlm.nih.gov/mesh/68020493), so that all articles indexed as Autobiography are automatically considered to be a type of Biography as well.

In summary, to deal with the fact that PTs may be overlapping and diverse, we feel that PT terms ought to be assigned to articles with probabilistic predictive scores estimating how well they fit each PT, i.e., how typical they are of each category.

## A PHILOSOPHICAL ISSUE

Should each article be indexed as an independent entity, or is the relevant unit the underlying study that the article is reporting on? The answer may appear to be obvious, but in practice, we have found that NLM sometimes uses the underlying study for indexing. The case of Adaptive

Clinical Trial[PT] is instructive, since these trials are very complex and give rise to a large number of articles that each report on subsets or specific aspects of one overall trial. We found that articles reporting single-arm trials within a larger adaptive trial were often indexed as Adaptive Clinical Trial even though the article itself did not have adaptive design (20). We propose that each article should explicitly be tagged according to its own design. However, separately, since for evidence synthesis, the STUDY (not the REPORT) is the unit of analysis, efforts should also be made to identify and aggregate all publications that arise from each trial (21-23). PubMed already maintains structured links between some related publication types, e.g., errata, retractions, and commentaries (https://www.nlm.nih.gov/bsd/policy/errata.html), demonstrating existing precedent for cross-article linkage when publications refer to the same underlying work.

**A UNIFIED HIERARCHY FOR PUBLICATION TYPES AND STUDY DESIGNS**

Does it matter on which tree a particular MeSH term resides? Or where it resides relative to others in the hierarchy? Yes, because these are the basis for automated query expansion as performed by PubMed. As mentioned above, a search on “central nervous system” will also automatically retrieve articles dealing with more specific terms such as “brain” and “spinal cord”. However, the existing NLM schema has some anomalies and limitations which need to be corrected to handle PTs and study designs properly.

The fact that topical MeSH terms and PT terms reside in separate trees causes confusion when users try to retrieve articles of a given type. For example, logically, Observational Study[PT] should be assigned to any article that satisfies the MeSH database definition: “A work that reports on the results of a clinical study in which participants may receive diagnostic, therapeutic, or other types of interventions, but the investigator does not assign participants to specific interventions (as in an interventional study)" (https://www.ncbi.nlm.nih.gov/mesh/68064888). There are 189,845 articles indexed as Observational Study[PT] as of January 21, 2026, but these actually do NOT include the vast majority of articles that have observational study designs, such as cohort studies, case-control studies, or cross-sectional studies. In fact, 367,604 articles are indexed as Cohort Studies[MH] alone, of which only 15,961 articles are also tagged as Observational Study[PT]. Because Observational Study[PT] and Cohort Studies[MeSH] reside on different trees, automatic expansion does not capture all observational study designs when carrying out PubMed searches.

A second problem is that some MeSH hierarchy assignments are arguably incorrect. For example, Retrospective Studies and Prospective Studies are both placed in the hierarchy under both Cohort Studies and Case-Control Studies, which is doubly wrong since many articles with retrospective or prospective designs are neither cohort nor case-control studies.

A third issue is that many Publication Types imply particular study designs; for example, Randomized Controlled Trial (RCT) is listed as a Publication Type but is associated with definite design features, some of which are represented by MeSH terms that reside on different trees (e.g., Random Allocation, often Double-Blind Method, often Placebos).

Thus, it would be desirable to merge publication types and study design-related MeSH terms into a single hierarchy. To accomplish this, we recently performed a three-step procedure (24). First, we calculated the pairwise similarity for all PTs and study designs. Second, hierarchical clustering was employed to group the most similar PTs together into 13 low-level categories and 5 broader categories. This information, combined with the existing NLM hierarchy (with corrections), allowed us to place all PTs and study designs into a single tree structure (24). This unified rubric and hierarchy permits more regular and appropriate expansion of PTs and study designs for indexing and retrieval.

**HOW ACCURATE IS PT INDEXING?**

Given that NLM curators were professionals who followed written guidelines, one would expect overall accuracy of indexing decisions to be quite high, especially for important and common publication types such as Randomized Controlled Trial. Indeed, our earlier analysis suggested that manual NLM indexers missed only ~3% of RCT articles and erroneously identified only ~5% of articles as RCT that were not (18). On the other hand, the annotation guidelines for NLM curators have not been made public, and we are not aware of any publications that have reported internal accuracy or consistency (e.g., agreement between independent annotators) of NLM indexing decisions of either topical MeSH terms or publication types and study designs. Publications that have manually examined highly specific MeSH terms and PTs have found much higher error rates, both with manual indexing and automated indexing methods. For example, 8% of articles indexed with the MeSH term *Malus* (the genus of the apple fruit) referred instead to other types of apples (acronyms, brand names, etc.) (25). Of articles indexed with the MeSH term Overdiagnosis (introduced in 2021), 26.8% were not actually discussing overdiagnosis, but rather different concepts such as misdiagnosis, false positive tests, or overtreatment (26). The MeSH term Papilledema was wrongly applied in 41% of case reports examined (27). We recently found that 46.7% of articles indexed as Adaptive Clinical Trial[PT] were mis-assigned insofar as they did not actually employ that design (20). Articles reporting survival of dental prostheses missed relevant MeSH terms in 30% of those examined (28). In a study of articles indexed using the MTIA algorithm, 47% had inadequacies in the indexing which could impact their retrieval, including inappropriate MeSH assigned; more general MeSH assigned while a more precise MeSH is available; or a significant concept not represented in the indexing at all (29). To our knowledge, the indexing accuracy of the most recently implemented MTIX algorithm has not yet been independently characterized.

Errors can be of two main types: **mis-assignment**, meaning an article was assigned one or more terms that were inappropriate, and **lack of coverage**, i.e. missing articles that should have been identified. Mis-assignment by indexers may occur when they take the word of the authors regarding their study design, e.g., if authors state explicitly that they are reporting a case-control study, especially in the title, there may be a tendency to accept that without undue scrutiny. This may also arise from variation in how authors interpret and use methodological terminology, from systematic differences in terminology across fields (e.g., computer science papers often refer to scoping reviews as systematic reviews), or from the inherently fluid nature of newly evolving study designs that lack rigid, universally agreed-upon definitions. Lack of coverage has not been well documented in the literature, but our own ongoing studies indicate that, at least for several important PTs (cohort studies, case-control studies, cross-sectional studies, and case reports), there is substantial lack of coverage of applicable articles (unpublished observations; see also (30)).

## EVALUATION OF AUTOMATED INDEXING SYSTEMS

Automated PT tagging systems have employed SVM-based platforms (as in the earlier RCT Tagger (18) and Multi-Tagger model (31)), transformer-based platforms (5, 30, 32), and LLM-based platforms (33). To date, nearly all large-scale evaluations of these systems have been both trained and evaluated utilizing the manual assignments of NLM curators as gold standards. However, the gold standards are far from perfect: Mis-assignment and lack of coverage cast doubt on the sole use of NLM indexing, either for training automated systems or for evaluating their outputs. An absolutely perfect indexing system would not show 100% accuracy against the gold standard! Rather, any correct scoring would be counted as error if NLM had not made the same decision. Also, some evaluations consider only the PT attached to an article's XML metadata and fail to take into account the hierarchy of more general terms – so that if an article is indexed as Cohort Studies in their XML metadata (but not explicitly as Observational Study), a system that marks the article as Observational Study would be counted as incorrect.

To evaluate an automated system most accurately, we propose that the most clinically important PTs should have new, independent, manually curated sets of articles created by a team of annotators following written definitions and scoring notes, and following the usual annotation guidelines (independent scoring by two annotators with reconciliation by a third when necessary). These can be used as gold standards for evaluation moving forward.

Actually, evaluating an automated PT system is not simply a matter of scoring correct vs. incorrect predictions, for several reasons:

First, when probabilistic scoring is conducted, there is initially no yes/no binary prediction, but rather an output score that ranges between 0 and 1. So the most important evaluation is the **calibration** of predictive scores (34). To do this, one divides the set of articles into bins according to their predictive scores between 0 and 1 (e.g., 0-0.1, 0.1-0.2, 0.2-0.3,…0.9-1.0), and checks the proportion of articles that received NLM indexing within each bin. If a model is well-calibrated, i.e,, has low expected calibration error (ECE) (34), a set of articles that have predictive scores of (say) 0.5 will have NLM indexing for 50% of those articles.

In practice, most users will want to convert the predictive scores into binary yes/no assignments, however, different users may wish to set very different thresholds depending on their need to optimize either recall or precision. For example, when screening articles to identify RCTs for writing systematic reviews, their goal is not to automatically find all RCTs, but rather to automatically discard all articles that are reliably **not** RCTs. To achieve recall as close to 100% as possible, setting a very low threshold of 0.01 ensures that one can confidently discard all articles having predictive scores below 0.01 without losing any appreciable RCT articles (35, 36). An increasingly popular strategy, especially for identification of RCT articles, is to employ the predictions of an automated RCT indexer as a "first screener" coupled with a second person acting as second screener (37).

Second, not all errors are equally important or egregious. For example, mis-assigning an Autobiography as a Biography might be regarded as a minor error -- and arguably not even as an error at all, since as discussed above, according to the NLM schema all Autobiography articles are automatically considered a type of Biography as well.

## INDEXING NEW, RARE AND NONSTANDARD PTs

Indexing schemes need to evolve over time as new types of publications appear (e.g. network meta-analysis or scoping review) and as gaps in current indexing terms become more glaring (e.g. there is no current indexing for Diagnostic Test Accuracy despite its popularity in evidence syntheses) (38). This is particularly evident for pre-clinical and translational research, including in vivo, in vitro, and in silico studies, which are foundational to hypothesis generation, mechanism elucidation, and early therapeutic development (39), yet remain entirely unrepresented in current indexing schemes. Despite their central role in the biomedical research pipeline, such studies are often treated as second-class citizens in indexing systems that privilege direct human applicability. Recent efforts, such as the GoldHamster dataset and associated models, have begun to develop automated approaches for identifying and characterizing animal-based study designs, highlighting both the feasibility of such indexing and the unmet need it addresses (33, 40). More broadly, the absence of systematic indexing for these publication types not only limits discoverability within pre-clinical domains, but also hampers integrative evidence synthesis across translational medicine. Addressing these gaps will require indexing strategies that explicitly accommodate rare, emerging, and non-human study designs, rather than implicitly marginalizing them through omission. Our unified hierarchy (24) can be readily updated to accommodate new PTs as they arise.

## DISCUSSION AND CONCLUSIONS

We hope we have demonstrated convincingly that the indexing of publication types and study designs (collectively, PTs) has distinct goals and requirements from the indexing of topical

MeSH terms. One should assign a PT, (say) Randomized Controlled Trial, to an article only if that article satisfies the specific requirements of experimental design, not if it merely discusses the topic of RCTs. One should index **all** PTs that apply to a given article, in a consistent, comprehensive manner, and using probabilistic prediction scores. Only then can evidence synthesis groups have enough confidence in the indexing that they can use PT indexing to retrieve only articles having desired study design(s), or alternatively may seek to remove the bulk of articles that lack the desired study design(s). Either way, the use of PT indexing should save substantial effort during triage. Casual users and science-of-science investigators alike will benefit from having better indexing of publication types and study designs.

Another benefit of an automated PT indexing system is that it is not restricted to PubMed alone. If users are allowed to provide article records (or even full-text article PDFs) as input and receive scores in real time, this strategy should offer consistent indexing across biomedical articles found in other bibliographic databases such as Scopus, Web of Science, EMBASE, and perhaps OpenAlex (41). Currently, transferring search strategies across databases (e.g., PubMed to Scopus), as is often done in systematic reviews, typically requires expert librarian involvement or the use of specialized translation tools such as Polyglot (42). An automated PT indexing system could mitigate this dependence by offering a unified, database-agnostic approach to study design indexing. Such a system could even be applied to preprints and unpublished manuscripts.

PT indexing systems currently have some definite limitations. For example, they all employ features derived from article title, abstract, and other metadata (5, 18, 30-32). This limits their ability to accurately assign PTs to articles missing abstracts, which have reduced textual data. In our preliminary studies using the transformer model (32), articles lacking abstracts (15% of articles overall) exhibited reduced indexing performance by 7% relative to those having abstracts (unpublished observations). Even in articles that have abstracts, their study designs may not be described in the abstract but require examination of the full-text (15, 43). Preliminary studies have demonstrated that extracting features directly from the full-text can improve performance above the level obtained by considering title, abstract, and metadata only (30). Further improvements may be expected using LLMs to process the full-text, together with article section detection to identify the most important sections for PT classification (especially, the Methods section) (30).

**Conflict of Interest statement**

The authors declare that they have no competing interests.

**Funding**

This work was supported by the National Library of Medicine at the National Institutes of Health [1R01LM014292-01 to N.R.S.]. Funder had no influence on the study, its design, or its publication.

**Data Availability**

Not applicable.

**Author Contributions**

NRS: Conceptualization, Funding acquisition, Methodology, Supervision, Writing - original draft, Writing – review & editing.

JDM: Methodology, Software, Investigation, Validation, Writing – review & editing.

AWH: Methodology, Formal analysis, Investigation, Writing – review & editing.

HK: Conceptualization, Methodology, Supervision, Writing – review & editing.

JS: Investigation, Supervision, Writing – review & editing.

**Acknowledgments**

An earlier version of this paper has been deposited to arXiv (44).